\documentclass{mem}
\usepackage{natbib}\usepackage{txfonts}\usepackage{balance}
\usepackage{graphicx}
\usepackage[a4paper]{hyperref}
\idline{75}{282}
\begin{document}
\def\teff{$T\rm_{eff }$}
\def\kms{$\mathrm {km s}^{-1}$}

\title{
B, V, I photometry of the complete sample of 23 Cepheids in the field of NGC 1866
}

   \subtitle{}

\author{
I. \,Musella\inst{1} 
\and V. \,Ripepi\inst{1}
\and E. \,Brocato\inst{2}
\and V. \,Castellani\inst{3,4}
\and F. \,Caputo\inst{3}
\and M. \,Del Principe\inst{2}
\and M. \,Marconi\inst{1}
\and A.M. \,Piersimoni\inst{2}
\and G. \,Raimondo\inst{2}
\and P.B. \,Stetson\inst{5}
\and A.R. \,Walker\inst{6}
}

  \offprints{I. Musella}

\institute{
Istituto Nazionale di Astrofisica --
Osservatorio Astronomico di Capodimonte, Vicolo Moiariello 16,
I-80131 Napoli, Italy; \email{ilaria@na.astro.it}
\and
Istituto Nazionale di Astrofisica --
Osservatorio Astronomico di Collurania, Via M. Maggini,
I-64100 Teramo, Italy 
\and
Istituto Nazionale di Astrofisica --
Osservatorio Astronomico di Monteporzio, Via di Frascati 33,
I-00040 Monte Porzio Catone, Italy 
\and
INFN - Sezione di Ferrara
\and
DAO --
Victoria, Canada
\and
Cerro Tololo Inter-American Observatory, Casilla 603, La Serena, Chile
}

\authorrunning{Musella}

\titlerunning{Cepheids in NGC 1866}

\abstract{
We present the result of $BVI$ photometry, obtained by using FORS@VLT,
of the Cepheids present in the field of the Large Magellanic Cloud
cluster NGC 1866. We found the 22 known variables plus an additional
new Cepheid located about 10' from the cluster center. The accuracy of
the photometry allowed us to derive $B$, $V$ and $I$ mean magnitudes
with an uncertainty lower than 0.02 mag for 22 out of
the 23 objects, with the exception of only one Cepheid (WS9) which
presents a noisy light curve due to the probable occurrence of image
blending. As a result, we provide accurate observational data for a
substantial sample of variables all lying at the same distance and
with a common original composition. The resulting period-luminosity
relations are presented and briefly discussed.

}
\maketitle{}

\section{Introduction}

Classical Cepheids play a fundamental role in the extragalactic
distance scale. Empirical calibrations of the Period-Luminosity (PL)
and Period-Luminosity-Color (PLC) relations are generally based on
field Cepheids involving the uncertainties due to the spread in
distance, metallicity and reddening. On the other hand, any
theoretical scenario for pulsational models, to provide a robust
support to empirical calibrations, needs to be confirmed through the
comparison with observational data as given by suitable samples of
well-observed Cepheids. Therefore, classical Cepheid members of young
stellar cluster (at the same distance and with a common chemical
composition and age) offer the opportunity to investigate the
uncertainties affecting both empirical and theoretical estimates
concerning their luminosity, color and periods.  The young Large
Magellanic Cloud (LMC) globular cluster NGC1866, already known to host
an exceptionally rich sample of Cepheids, as given by at least 22
variables (\citealt{WS93} and references therein) represents an ideal
candidate. However, in a recent paper, \cite{B04} have investigated
this sample, reaching the unpleasant conclusion that only 4-6 of the
known Cepheids have light curves accurate enough to allow a meaningful
determination of their luminosities and colors. There remain large
errors for the variables in the central crowded region. In this
context, we took advantage of assigned observing time at the ESO Very
Large Telescope to perform a new photometric investigation of the
cluster field with the aim to obtain suitable sampling of the light
curves of all the Cepheids members of the cluster.

\begin{figure}[]
\resizebox{\hsize}{!}{\includegraphics[clip=true]{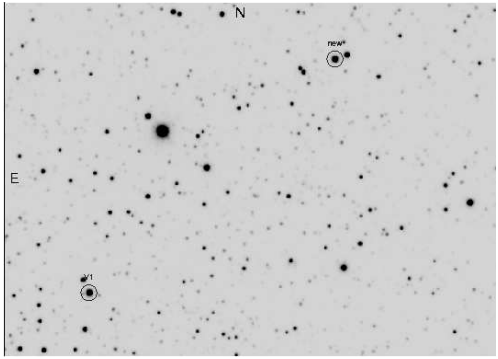}}
{\caption{\footnotesize Identification map for the newly discovered
Cepheid in the field of NGC 1866.
}}
\label{fig:map_new}

\vspace{-0.5truecm}

\end{figure}

\section{Observation and data reduction.}

Observations have been carried out by means of the FORS1@VLT
instrument in imaging mode. The detector was a 2048x2048 Tektronix CCD
with pixel size $24\mu \times 24\mu$. Projected on the sky, the pixel
size is 0.2 arcsec/pixel for a total field of view of $6.8^{'} \times
6.8^{'}$.  We observed one field centered on NGC1866 and in total we
got 69 images in $B$, 90 in $V$ and 62 in $I$ with an exposure time of
60 sec for each image in all three bands and a median seeing better
than $0.7^{''}$. Moreover, due to the periods of 3-4 days for the
Cepheids in NGC1866, we require a time coverage of about three months
in order to find reasonable periods for the variables.  In order to
pre-reduce the data we followed the standard procedure by de-biasing
and flat-fielding the images. Photometry has been carried out by
means of DAOPHOT/ALLFRAME packages (\citealt{S87,S92}) which couple
excellent precision with high degree of automation. Special care has been
taken in deriving an accurate PSF for each image because of the high
degree of crowding not only in the central regions of the cluster but
also in its outskirts. The master list for ALLFRAME was built by
running DAOPHOT/ALLSTAR on an image which is the combination of the
best 10 $B$, 10 $V$ and 10 $I$ frames. The threshold of detection has
been selected in order to avoid the presence of a high number
of spurious stars.  The calibration to the standard system was
obtained by using the Stetson local standards (\citealt{S00},
http://cadcwww.hia.nrc.ca/standards) and the final error is lower than
0.02 mag in all the bands.

\subsection{NGC 1866 Cepheids} 

\begin{figure}[t!]
\resizebox{\hsize}{!}{\includegraphics[clip=true]{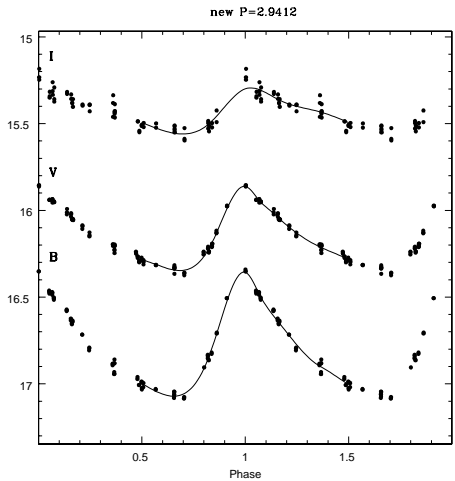}}

\vspace{-0.4truecm}

\caption{\footnotesize
The $BVI$ light curves for the new Cepheid found in NGC 1866.}
\label{fig:new}

\vspace{-0.5truecm}

\end{figure}

\begin{figure}[t!]
\resizebox{\hsize}{!}{\includegraphics[clip=true]{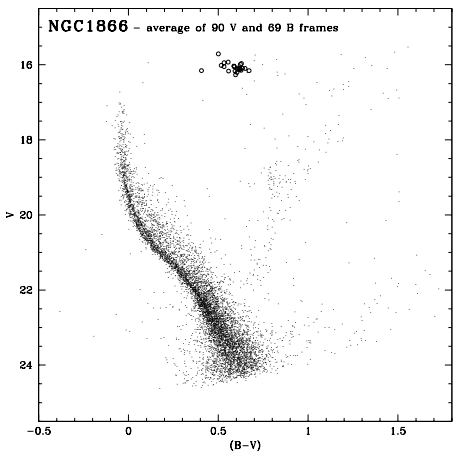}}

\vspace{-0.4truecm}

\caption{\footnotesize
The color magnitude
diagram of NGC 1866 as derived in this work. The locations of the
22 well observed Cepheids studied in this work are indicated as
open circles.}
\label{fig:cm}

\vspace{-0.5truecm}

\end{figure}

\begin{figure*}[t!]
\resizebox{\hsize}{!}{\includegraphics[clip=true]{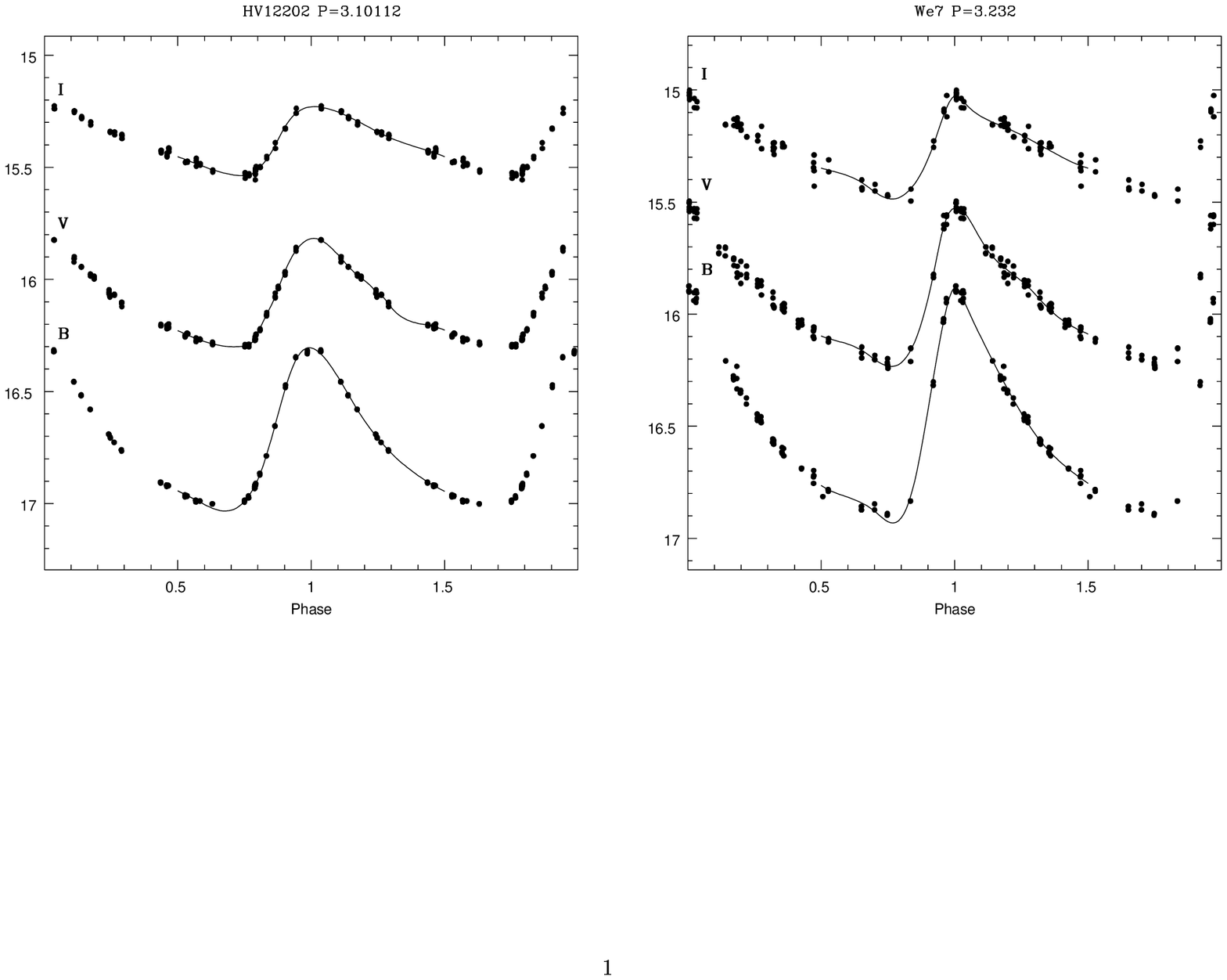}}

\vspace{-0.3truecm}

\caption{\footnotesize
The light curves of the Cepheids HV12202 and We7, located in a
peripheral region and in the most crowded central region,
respectively.}
\label{fig:LC}

\vspace{-0.5truecm}

\end{figure*}

\begin{figure}[t!]
\resizebox{0.8\hsize}{!}{\includegraphics[clip=true]{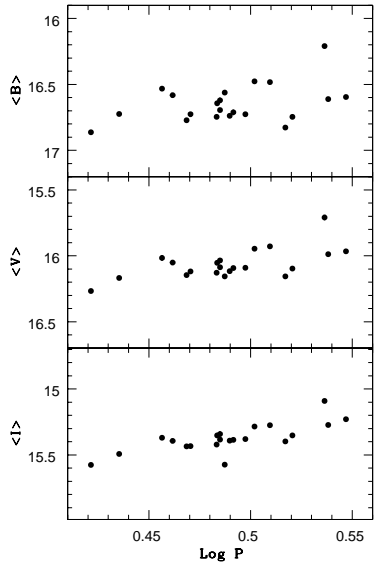}}

\vspace{-0.5truecm}

\caption{\footnotesize
The PL relationships, in the $B$, $V$ and $I$ bands,
for the sample of 22 Cepheids.}
\label{fig:pl}

\vspace{-0.5truecm}

\end{figure}

As a result, we secured photometric data for all the 22 known
Cepheids, plus a newly discovered Cepheid whose identification map is
reported in Fig. \ref{fig:map_new}.  The light curve for the new
variable is reported in Fig. \ref{fig:new}. Suspicion about the cluster
membership of this new Cepheid may arise due to its location in the
extreme periphery of the cluster, but the position in the CMD
(Fig. \ref{fig:cm}) and the pulsational properties of this variable
appear to be in close agreement with the other cluster Cepheids,
supporting the idea that this star is in the same evolutionary state
as the other cluster variables.  The CMD in Fig. \ref{fig:cm},
obtained from the average of 90 $V$ and 60 $B$ frames, has a main
sequence very well defined and shows the high quality of the
photometric data. In particular, we have obtained very accurate light
curves for 22 out of 23 Cepheids present in this field, including
those located in the central crowded region (see the examples in
Fig. \ref{fig:LC}), and evaluated the amplitudes and colors with an
error lower than 0.02 mag. In particular, a comparison with the mean
magnitudes given in \cite{B04} for the six peripheral well-studied
Cepheids shows an agreement within this error.  The only object with a
noisy light curve (due to the probable occurrence of image
blending) is WS9, which will not be included in the following
discussion.
Fig. \ref{fig:cm} shows the positions of the Cepheids (open circles) in
the CMD. Not surprisingly, one finds that when dealing with accurate
magnitudes and colors all the Cepheids appear concentrated in a
restricted region of the diagram, corresponding to the tip of the
 He-burning giant blue loop.The only exception is HV 1204
(see \citealt{B04}), which appears slightly bluer and more luminous.
Fig. \ref{fig:pl} shows the $BVI$ PL relations for the whole set of
Cepheids. One may easily detect the occurrence of two First Overtone
(FO) pulsators (with $\log{P}<0.44$). Moreover, since the large
majority of Cepheids can be regarded as bona fide cluster members,
data in figure \ref{fig:pl} give for the first time, to our knowledge,
a robust evidence of the scatter in magnitude for each given period
from a sample of stars with the same age and original chemical
composition.

\section{Discussion and conclusions}

\begin{figure}[t!]
\resizebox{\hsize}{!}{\includegraphics[clip=true]{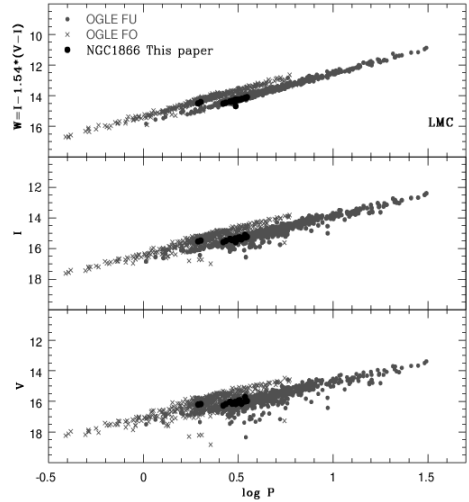}}

\vspace{-0.3truecm}

\caption{\footnotesize
The $V$, $I$ and Wesenheit PL relations for the Cepheids in NGC1866
(black dots) as compared with the LMC Cepheid sample by OGLE (gray
dots represent fundamental pulsators and gray crosses the first
overtone ones).}
\label{fig:ogle}

\vspace{-0.5truecm}

\end{figure}

As a final point, Fig. \ref{fig:ogle} compares the $V$, $I$ and
Wesenheit P-L relations of NGC1866 Cepheids with the same relation
for the large LMC OGLE sample (including fundamental and first
overtone pulsators). As a result, one finds that the two sample agree
with each other very well, showing that NGC1866 Cepheids have to be
regarded as a bona fide sample of LMC Cepheids.  Moreover, one finds
that, in the Period-Magnitude plane, NGC1866 Cepheids appear slightly
less luminous than the mean LMC sequence for both fundamental (FU) and
First Overtone (FO) pulsators. This could be interpreted as 
evidence that NGC1866 is slightly more distant than the main body of
LMC Cepheids. If this is the case, the NGC1866 distance modulus
appears larger by about 0.1 mag. This is only a preliminary
discussion, as all the data and a more complete study will be
presented in a future paper (Ripepi et al. in preparation). The
large and homogeneous sample of Cepheids belonging to the star cluster
NGC 1866 will allow us to improve the knowledge of the distance of
NGC1866 and therefore of the LMC and to test the predictive capabilities
of current theoretical models to provide independent constraints on
the physical and numerical assumptions adopted in the pulsational
models.

\begin{acknowledgements}
This paper is based on observations made with the ESO/VLT.

Financial support for this work was provided by MIUR-Cofin 2003,
under the scientific project ``Continuity and Discontinuity in the
Milky Way Formation'' (P.I.: Raffaele Gratton). This project made use
of computational resources granted by the Consorzio di Ricerca del
Gran Sasso according to the Progetto 6 'Calcolo Evoluto e sue
Applicazioni (RSV6)' - Cluster C11/B.
\end{acknowledgements}

\bibliographystyle{aa}

\end{document}